\documentclass[conference]{IEEEtran}
\IEEEoverridecommandlockouts
\usepackage{amsmath,amssymb,amsfonts}
\usepackage[utf8]{inputenc}
\usepackage{algpseudocode}
\usepackage{cite}
\usepackage{algorithm}
\usepackage{graphicx}
\usepackage{textcomp}
\usepackage{url}
\usepackage[caption=false]{subfig}
\usepackage[justification=centering]{caption}
\usepackage{epsfig}
\usepackage{float}
\usepackage{color}
\usepackage{balance}
\usepackage{bbm}
\usepackage{textcomp}
\usepackage{enumitem}
\usepackage{mathtools}
\ifCLASSINFOpdf
\else
\fi
\newcommand{\linebreakand}{%
  \end{@IEEEauthorhalign}
  \hfill\mbox{}\par
  \mbox{}\hfill\begin{@IEEEauthorhalign}
}

\begin{document}

\title{ Reconfigurable Intelligent Surface Aided
Mobile Edge Computing over Intermittent mmWave Links\thanks{The work of Airod, Di Lorenzo and Calvanese Strinati has been partially funded by the H2020 project RISE-6G no. 101017011.}}

\author{Fatima Ezzahra Airod$^1$, Mattia Merluzzi$^1$, Paolo Di Lorenzo$^{2,3}$, Emilio Calvanese Strinati$^1$\\
$^1$CEA-Leti, Université Grenoble Alpes, F-38000 Grenoble, France\\
$^2$DIET department, Sapienza University of Rome, Italy, $^3$CNIT, Parma, Italy\\
email:\{fatima-ezzahra.airod, mattia.merluzzi, emilio.calvanese-strinati\}@cea.fr,\{paolo.dilorenzo\}@uniroma1.it\vspace{-.35 cm}}
\maketitle
\begin{abstract}
The advent of Reconfigurable Intelligent Surfaces (RISs) in wireless communication networks unlocks the way to support high frequency radio access (e.g. in millimeter wave) while overcoming their sensitivity to the presence of deep fading and blockages. In support of this vision, this work exhibits the forward-looking perception of using RIS to enhance the connectivity of the communication links in edge computing scenarios, to support computation offloading services.
We consider a multi-user MIMO system, and we formulate a long-term optimization problem aiming to ensure a bounded end-to-end delay with the minimum users' average transmit power, by jointly selecting uplink user precoding, RIS reflectivity parameters, and computation resources at a mobile edge host. Thanks to the marriage of Lyapunov stochastic  optimization, projected gradient techniques and convex optimization, the problem is efficiently solved in a per-slot basis, requiring only the observation of instantaneous realizations of time-varying radio channels and task arrivals, and that of communication and computing buffers. Numerical simulations show the effectiveness of our method and the benefits of the RIS, in striking the best trade-off between power consumption and delay for different blocking conditions, also when different levels of channel knowledge are assumed.
\end{abstract}


%
\IEEEpeerreviewmaketitle
\section{Introduction}
The advent of the sixth generation of mobile communication systems (6G) unveils several ambitions that span from the support of new services, to completely new key performance indicators. 
Indeed, we are facing an unprecedented revolution of applications, such as immersive virtual reality, connected autonomous systems, and the industrial Internet of Things, all verticals that require real time data transmission and processing \cite{1}. Of course, to be effective, these services require to be enabled with new levels of dependability, reliability and sustainability. 
From a radio access perspective, the adoption of higher frequency such as millimeter wave (mmWave) and TeraHertz (THz) bands certainly enhances radio access network capacity, although at the price of a higher sensitivity to the presence of spatial blockages and, in general, to deep fading events that may hinder the aforementioned vision on performance \cite{mmwv1,mmwv2}. To this end, Reconfigurable Intelligent Surfaces (RISs) have recently emerged as a promising candidate to counteract the above mentioned issue, thanks to their ability to opportunistically shape the wireless propagation environment. 
More precisely, RISs are composed of scattering elements that can be adaptively configured to shape the incident wave through adjustable phase shifts, with the aim of improving system performance in specific locations in space and time \cite{4,5}.
Therefore, owing to their abilities of customizing time-varying wireless propagation environments, RISs mark   undeniably   the   dawn   of  the 6G era. 
Another key technological enabler of the 6G vision, already introduced in 5G, is Multi-access Edge Computing (MEC), which brings storage and computing resources close to end users, aiming to enable a new class of connect-compute services \cite{7}. As such, the interplay between RISs and MEC plays a key role in improving network performance, thanks to the double benefit of computation and communication aspects, to be tackled and optimized jointly. 
In this paper, we focus on computation offloading services, whose goal is to move the execution of computation demanding applications from resource-poor end devices to nearby Mobile Edge Hosts (MEHs), to enable energy efficient, low latency, and reliable processing \cite{14}. In particular, we investigate on the promising convergence of RISs and MEC, mainly focusing on a joint optimization of radio and computing resources, down to the wireless propagation environment properties. \\
\textbf{Related works.} Most of investigated works in the literature have been focused on addressing computation offloading upon appropriate wireless environments, tackling the joint optimization of communication and computation resources in 
MEC-enabled wireless networks \cite{7,11}. However, inevitably, moving towards higher frequency bands to cope with large data volumes is no more suitable for MEC systems due to the unpredictable and intermittent nature of wireless links. Indeed, blocking events may deteriorate the overall network performance. In line with this, the recent literature has involved the prominence of RISs to boost the performance of MEC systems in terms reliability \cite{13,14,15}. 
Nevertheless, to the best of our knowledge, a dynamic joint optimization of computing resources and RIS-aided MIMO radio parameters is lacking.\\
\textbf{Contribution.} In this work, we propose an algorithm aiming to dynamically configure RIS parameters, users' uplink precoding, and computation resources, with the goal of minimizing users' transmit power, with guaranteed finite E2E delay of the offloading service. Thanks to the theory of Lyapunov stochastic optimization, we are able to split a long-term problem into consecutive deterministic optimization problems, based on instantaneous observations of context parameters. 
The solution of the latter, from a radio perspective, builds on an alternating optimization strategy that couples a projected gradient step for the RIS parameters \cite{pgm}, and a water-filling solution for the users' precoding \cite{wtfil}. The computation resource allocation problem is also solved with low complexity.

\section{System Model}
In this work, we consider a dynamic system, in which a set $\mathcal{U}$ of $N$ users continuously generate data locally and offload them to an MEH through the wireless connection with an Access Point (AP). In such a scenario, context parameters such as wireless channels and data arrivals vary over time, thus calling for a dynamic cross-layer optimization involving users' precoding for uplink transmission, RIS reflectivity matrix, and computation resources at the MEH. Then, we consider time as organized in time slots $t=1,2,\ldots$ of equal duration $\tau$. Given a random variable $X$, we denote by $\overline{X}$ its long-term average:
\begin{equation}\label{long_term_avg}
    \overline{X}=\lim_{T\to\infty}\frac{1}{T}\sum\nolimits_{t=1}^T\mathbb{E}\{X(t)\}
\end{equation}

\subsection{Communication model}
\subsubsection{RIS reflectivity model}
An RIS can be modeled as an array of $M$ nearly passive elements, whose phases can be opportunistically tuned. Therefore, the RIS can be characterized by its reflectivity matrix that is represented, at time $t$, as $\mathbf{\Theta}(t)=\textrm{diag}\{r_1(t),...,r_M(t)\}$, where $r_i$, $i=1...,M$, are the complex reflection coefficients of each element, characterized by a fixed amplitude, which we assume to be unitary, and an adjustable phase shift $\theta_i$, therefore we have $r_i=\alpha_ie^{j\theta_i}$. In the sequel, we will denote by $\mathbf{r}(t)$ the vector $\mathbf{r}(t)=\{r_i(t)\}_{i=1}^M$.
\subsubsection{Channel model}
We consider a MIMO system, in which the AP is equipped with $N_a$ antennas, while each user is equipped with $K$ antennas. For each user $k$, the E2E channel matrix $\mathbf{H}_k(t)$ at time $t$ is composed of: i) a direct channel $\mathbf{H}_{k,d}(t)\in \mathbb{C}^{N_a\times K}$ between the user and the AP; ii) an indirect link, comprising the channel $\mathbf{H}_{{k},r}(t) \in \mathbb{C}^{{M}\times K}$ between the user and the RIS, and the channel $\mathbf{H}_{{r,a}}(t) \in \mathbb{C}^{N_a\times M}$ between the RIS and the AP. Also, since we consider blocking, we define $\beta_{k,a}(t)\in\{0,1\}$ ($\beta_{k,r}(t)\in\{0,1\}$), which equals $1$ if the direct (indirect) link experiences a blockage event. Then, the overall channel matrix of user $k$ can be written as follows \cite{simR} (we omit the index $t$ for ease of notation):
\begin{equation}\label{chan}
\normalsize
\mathbf{H}_k = (1-\beta_{k,a})\mathbf{H}_{k,d} +(1-\beta_{k,r})\mathbf{H}_{{r,a}}\Theta  \mathbf{H}_{k,r}.
\end{equation}
Note that, in the sequel, we will denote by $p_{k,a}$ ($p_{k,r}$) the blocking probability of the direct (indirect) link, i.e. the probability that $\beta_{k,a}$ ($\beta_{k,r}$) equals $1$, which can be computed as the expectation of $\beta_{k,a}$ ($\beta_{k,r}$). Uplink transmission is a fundamental phase of computation offloading services, foreseen to also increase future uplink traffic \cite{ulMEC}. Thus, let us now formalize the uplink parameters, whose optimization will be presented in the sequel. Letting $\mathbf{Q}_k(t)\in\mathbb{C}^{K\times K}$ be the transmit covariance matrix of user $k$ at time $t$, the experienced data rate reads as follows
\begin{equation}\label{capacity}
          R_{k}(t) = W_k \log_2\left|\mathbf{I}+\sigma_k^{-2}\mathbf{H}_k(t)\mathbf{Q}_k(t)\mathbf{H}_k^{H}(t)\right|
\end{equation}
where $W_k$ represents the bandwidth assigned to  user $k$, and the noise power is $\sigma_k^2=N_0W_k$, with $N_0$ the noise power spectral density. Here, we assume  that users are served thorough orthogonal channels with a frequency division multiplexing. 
Obviously, besides the current channel state conditions, the data rate depends on the user transmit covariance and the RIS parameters, which we will jointly optimize in the sequel.
\subsection{Queuing Model and delay}
Computation offloading services generally entail three phases, along with their respective delays: i) uplink communication buffering and transmission of input data; ii) computation buffering and computation; iii) downlink communication buffering and transmission of results. In this work, we consider the first two delays, although considering the last one would not substantially change the system model, as presented in \cite{14}. Therefore, we model the E2E delay thorough a  queueing system, comprising an uplink communication buffer $B_{l,k}(t)$, and a computation buffer $B_{r,k}(t)$ for each user $k$.\\
\textbf{Communication buffer}: The uplink buffer of each user $k$ is fed by new arrivals $A_{k}(t)$ at time $t$, and drained by transmitting bits over the wireless interface at rate $R_k(t)$ (cf. \eqref{capacity}). Given a slot of duration $\tau$, the queue evolves as: 
\begin{equation}\label{queue_evolution1}
\normalsize 
  B_{l,k}(t+1)=\max\left(0,B_{l,k}(t)-\tau R_{k}(t)\right) +A_k(t)
\end{equation}
\textbf{Computation buffer}: Assuming that all computation tasks are offloaded to the MEH, we consider a computation queue for each UE, which is fed by the its arriving data in uplink, and drained by the computation performed at the MEH. We assume a linear relation between the number of transmitted bits and the CPU cycles. 
Then, denoting by $J_{k}$ the number of CPU cycles per bit, the remote computation queue evolves as:
\begin{align}\label{queue_evolution2}
\normalsize
  B_{r,k}(t+1)=&\max\left(0,B_{r,k}(t)-\tau f_{k}(t)/J_{k}\right)\nonumber\\    
              &+\min(B_{l,k}(t),\tau R_{k}(t)),
\end{align}
where $f_k(t)$ represents the amount of resources (in CPU cycles/s) allocated to user $k$ during time slot $t$. Due to Little's law, the average E2E delay experienced by each device is proportional to the sum queue length \cite{7}:  $\overline{D_k}=\tau\frac{\overline{B_{l,k}}+\overline{B_{r,k}}}{\overline{A_k}}$. 
\section{Problem formulation}
In this paper,  we jointly optimize users' uplink covariance matrix, RIS parameters, and computation  resources  at  the  MEH, to minimize the users' transmit power under queue stability constraints. 
The problem can be formulated as follows:
 \begin{align}\label{lg_eq}
 \small
&\hspace{- 1.5 cm}\min_{\{\mathbf{Q}_{k}(t)\}_k, {\mathbf{r}(t)}, \{f_{k}(t)\}_k} \quad  \sum\nolimits_{k\in\mathcal{U}}\overline{\textrm{Tr}(\mathbf{Q}_{k})}\\
\textrm{subject to} \; &(a)\;  \overline{B_{l,k}} <\infty,\quad \forall k\;\; (b)\;  \overline{B_{r,k}} <\infty,\quad \forall k\nonumber \\
    &(c)\;  \mathbf{Q}_{k}(t)\succcurlyeq 0,\quad \forall k\;\, (d)\;  \textrm{Tr}(\mathbf{Q}_{k}(t))\leq P_k^{\max},\quad \forall k\nonumber\\
    &(e)\; |r_i(t)|=1,\; \forall i\quad\, (f)\; f_k(t)\geq 0,\quad \forall k\nonumber\\
    &(g)\; \sum\nolimits_{k\in\mathcal{U}} f_{k}(t)\leq f_{\max}.\nonumber
\end{align}  
The constraints of \eqref{lg_eq} have the following meaning: $(a)$-$(b)$ the local and remote queues of each user are stable; $(c)$ the transmit covariance matrix of each user is semidefinite positive; $(d)$ the uplink transmit power of each user is lower than a maximum value $P_k^{\max}$; $(e)$ the RIS reflectivity entries are complex exponential; $f)$ the CPU cycle frequency allocated to each user by the MEH is non-negative; $(g)$ The sum all CPU cycle frequencies assigned to each user is at most equal to the MEH CPU maximum frequency $f_{\max}$. Problem \eqref{lg_eq} is a priori very complex to solve, as it involves time averages performed on variables whose statistics are supposed to be unknown. To solve it in an efficient way, we leverage on Lyapunov stochastic optimization \cite{neely10}, which allows us to define a sequence of deterministic problems, based on instantaneous observations of context parameters.
In particular, following \cite{neely10}, and defining the vector
$\mathbf{b}(t)=[\{B_{l,k}(t)\}_k,\{B_{r,k}(t)\}_k]$, we can write the \textit{Lyapunov function} as $L(\mathbf{b}(t))=\frac{1}{2}\sum\nolimits_{k\in\mathcal{U}}[B_{l,k}^2(t)+B_{r,k}^2(t)]$ \cite{neely10},
which is a measure of the overall congestion state of the system. Our aim is to drive the network towards stability, with the minimum transmit power. To this end, as in \cite{neely10}, let us define first the \textit{drift-plus-penalty} (DPP) function $ \Delta_{p}(t)\!=\!\mathbb{E}\{\!L(\mathbf{b}(t+1))-L(\mathbf{b}(t))\!+\!V\sum_{k\in\mathcal{U}}\textrm{tr}(\mathbf{Q}_k(t))|\mathbf{b}(t)\!\},$
which is the one slot conditional expected change of the Lyapunov function, with a penalty factor, weighted by a parameter $V$, used to trade-off users' transmit powers and queue backlogs, thus shaping the desired trade-off between power consumption and E2E delay. Interestingly, queues' stability ($(a)$-$(b)$ in \eqref{lg_eq}) is guaranteed if the DPP is bounded by a finite constant for all $t$ \cite{neely10}. As in \cite{neely10}, we now proceed by minimizing a suitable upper bound of the DPP. The upper bound, whose derivations are omitted due to the lack of space (see, e.g., \cite{neely10}) reads as
\begin{align}
&\Delta_p(t)\leq C+ \mathbb{E}\big\{\sum\nolimits_{k\in\mathcal{U}}[\left(B_{r,k}(t)-B_{l,k}(t)\right)\tau R_k(t)\nonumber \\
&\!+A_k(t)B_{l,k}(t)-\tau B_{r,k}(t)f_{k}(t)/J_k+V\textrm{tr}(\mathbf{Q}_k(t))]|\mathbf{b}(t)\big\},\nonumber
\end{align}
where $C$ is a positive constant, omitted due to the lack of space. By greedily minimizing this upper bound in each time slot (i.e. removing the expectation), queues' stability is guaranteed, as well as the asymptotic optimality of the solution as $V$ increases, with the cost of increased queue backlogs (i.e. higher E2E delay) \cite{neely10}. It is easy to show that the resulting problem can be split, in each time slot $t$, into a radio resource allocation sub-problem, including the optimization of user covariance matrices and RIS parameters, and a computation resource allocation sub-problem, to optimize the MEH CPU scheduling. The overall proposed dynamic resource allocation procedure is described in Algorithm \ref{alg:cap}, whose steps are described in the following. In particular, Sec. \ref{sec:radio} describes the implementation of step $1$, to optimize radio resources. The implementation of step $2$ follows in Sec. \ref{sec:comp_res}. 
\subsection{Radio resource allocation sub-problem}\label{sec:radio}
The radio resource allocation sub-problem (step $1$ of Algorithm \ref{alg:cap}) involves $\{\mathbf{Q}_k(t)\}_k$ and $\mathbf{r}(t)$, and is formulated as
 \begin{align}\label{slot_problem_radio}\small
     \min_{\{\mathbf{Q}_{k}(t)\}_k, {\mathbf{r}(t)}} \;& \sum\limits_{k\in \mathcal{U}} \left(V\textrm{Tr}(\mathbf{Q}_{k}(t)) -\tau  (B_{l,k}(t)-B_{r,k}(t))R_{k}(t)\right)\nonumber \\
   &\textrm{subject to} \quad (c)\textrm{-}(e)\;\textrm{of}\;\eqref{lg_eq}
\end{align}
Problem \eqref{slot_problem_radio} is non convex, due to the non linear equality constraint $(e)$. 
However, given the RIS parameters, the problem is convex and enjoys a simple water-filling solution \cite{wtfil}. Then, the solution of \eqref{slot_problem_radio} is built on an iterative optimization algorithm that alternatively optimizes \eqref{slot_problem_radio} with respect to the RIS phase shift using the projected gradient descent method (PGM), as in \cite{pgm}, and optimally updates the uplink covariance matrices of all users through the water-filling method \cite{wtfil}. Steps $1.1$ and $1.2$ are implemented as follows.
\subsubsection{RIS optimization step (Step $1.1$ of Algorithm \ref{alg:cap})}
The projected gradient step (step  $1.1$) with respect to $\mathbf{r}(t)$ comes with low complexity, as both the gradient and the projection can be written in closed form \cite[Eq. ($17$a)]{pgm}. However, differently from \cite{pgm}, we deal with a multi-user case. Nevertheless, thanks to the decoupling obtained through the Lyapunov optimization framework, in this case, the gradient is a weighted sum of different terms (corresponding to different users), where the weights include both communication and computation queues. This naturally introduces a scheduling of the RIS, which is therefore optimized to prioritize users with worse queueing states. 
The gradient with respect to $\mathbf{r}$ reads as follows \cite{pgm}: 
 \begin{align}\label{grad}\small
   &\nabla_{\mathbf{r}}f(\textbf{r},\left\{\textbf{Q}_k\right\}_k)=-\tau \sum\nolimits_{k\in\mathcal{U}} W_{k}(B_{l,k}-B_{r,k})\nonumber\\
   &\times \textrm{diag}\left(\mathbf{H}_{r,a}^H\left(I+\mathbf{Z}_k\mathbf{Q}_k\mathbf{Z}_k^H\right)^{-1}\mathbf{Z}_k\mathbf{Q}_k\overline{\mathbf{H}}_{k,R}^H\right), 
 \end{align}
where $\mathbf{Z}_k=\mathbf{H}_k/\sigma$, and $\overline{\mathbf{H}}_{k,R}=\mathbf{H}_{k,r}/\sigma$, and the operator $\textrm{diag}(L)$ saves the diagonal elements of an $N_L\times N_L$ matrix $L$ into a vector.  Finally, since $|r_i|=1$ must hold for all $i=1,\ldots M$, the projection onto the unit circle reads as \cite{pgm}:
\begin{equation}\label{proj}
    P_{\Theta}(r_i)= r_i/|r_i|, \qquad\forall i=1,\ldots,M.
\end{equation}
Step $1.1$ of Algorithm \ref{alg:cap} is implemented through \eqref{grad} and \eqref{proj}.
\subsubsection{Uplink covariances optimization (Algorithm \ref{alg:wordy})}
From (\ref{slot_problem_radio}), it can be easily observed that, once the RIS configuration is fixed, the problem with respect to $\{\mathbf{Q}_k(t)\}_k$ admits a low complexity solution. First of all, the problem is separable among the $N$ users. Moreover, for a generic user $k$, if $B_{l,k}(t)\leq B_{r,k}(t)$, both terms in \eqref{slot_problem_radio} are monotone non-decreasing functions of the user transmit power. Therefore, in this case, the optimal solution is $\mathbf{Q}_k^*(t)=\mathbf{0}_{K}$, i.e. user $k$ does not transmit (step $1.2.a$ of Algorithm \ref{alg:cap}). This holds true also in the case in which all links are blocked. 
Instead, for a generic user $k$ for which $B_{l,k}(t)> B_{r,k}(t)$ holds, the problem is convex and is similar to the one presented in \cite{wtfil}, thus admitting the water-filling procedure in Algorithm \ref{alg:wordy}. 
\subsection{Computation resource allocation sub-problem}\label{sec:comp_res}
The second sub-problem, necessary to implement step $2$ of Algorithm \ref{alg:cap}, is formulated as follows:
\begin{align}\label{freq}
\normalsize
       \max_{f_{k}(t)} \quad   &\sum\nolimits_{k\in\mathcal{U}} B_{r,k}(t)f_{k}(t)/J_k \\
   \textrm{subject to} \quad & a)\; 0\leq f_k\leq\min\left(f_{\max},B_{r,k}(t)J_k/\tau\right),\quad\forall k\;\nonumber\\
    &b)\; \sum\nolimits_{k\in\mathcal{U}} f_{k}(t)\leq f_{\max},\nonumber
\end{align}
where, for efficiency purposes, we added the constraint in $(a)$ that prevents each user to be allocated more frequency than the one needed to empty the remote queue. Problem (\ref{freq}) is linear, and the optimal frequencies can be iteratively found by assigning the whole available frequency to the user with the highest ratio $B_{r,k}(t)/J_k$. If this leaves available frequency, the left part is assigned to the subsequent users until draining the whole CPU power of the server $f_{\max}$, or serving all users \cite{14}.

\begin{algorithm}[hbt!]
\caption{Dynamic optimization of RIS-assisted MEC}\label{alg:cap}
\begin{algorithmic}
\small
\Require $V$, $\mathcal{U}=\{1,\ldots,N\}$ $N_{slots}$, $\tau$, $P_k^{max}$, $B_{l,k}(0)$, $B_{r,k}(0)$, $J_k$, $\forall k\in\mathcal{U}$, $f_{\max}$, 
\For{$t=1:N_{slots}$}
\Statex  \textbf {step 1}: Optimize $\{\mathbf{Q}_k(t)\}_k$ and $\mathbf{r}(t)$.
\For{$n=1:I_{\max}$}
\Statex  \textbf {step 1.1}: $\mathbf{r}^{n+1}=P_{\Theta}(r_{n}-\rho\nabla_{\mathbf{r}}f(\mathbf{r}^{n},\{\mathbf{Q}_{k}^{(n)}\}_k))$
\Statex  \textbf {step 1.2}:\For{$k=1,\ldots,N$}
\State \textbf{a:} If $B_{l,k}\leq B_{r,k}$, $\mathbf{Q}_{k}^{(n+1)}=\mathbf{0}_{K}$, else
\State \textbf{b:} Update optimal $\{\mathbf{Q}_{k}^{(n+1)}\}_k$ with Algorithm \ref{alg:wordy}
\EndFor

\EndFor

\Statex  \textbf{step 2}: Optimize  $\{f_k(t)\}_k$ as in Section \ref{sec:comp_res}\label{s1.2}
\Statex  \textbf{step 3}:  Compute $R_k(t),\forall k\in\mathcal{U}$ as in \eqref{capacity};\label{s1.3}
\Statex \textbf{step 4}:  Update $B_{l,k}$ and $B_{r,k}$ as in (\ref{queue_evolution1}) and (\ref{queue_evolution2}), respectively. 
\EndFor
\end{algorithmic}
\end{algorithm}


\begin{algorithm}
\caption{Uplink covariance optimization for user $k$ \cite{wtfil}}\label{alg:wordy}
\begin{algorithmic}
\small
\Statex \textbf{step 1:}  Compute  $\mathbf{H}_k^H\mathbf{H}_k= \mathbf{U} ^H \mathbf{\Sigma} \mathbf{U}$, with $\mathbf{\Sigma}$ a diagonal matrix with non-negative elements $\sigma_{i}, i=1,...,K$
    \Statex \textbf{step 2:} Check if $\sum_{i=1}^{K} \max \left(0,\frac{\tau W_{k}(B_{l,k}-B_{r,k})}{V}-\frac{1}{\sigma_{i}}\right)\leq P_k^{\max}$ holds. If yes, then let $\mu^{*}=0$ \\and $\lambda^{*}_{i}=\max\left(0,\frac{\tau W_{k}(B_{l,k}-B_{r,k})}{V}-\frac{1}{\sigma_{i}}\right)$, else,\\
    \textbf{step 3:} Take all $\sigma_{i},  i=1,\ldots,K$ in a decreasing order, i.e. as, $\sigma_{d(1)}>\sigma_{d(2)}>\ldots>\sigma_{d(K)}$.\\
     \textbf{step 4:} \\ Start with $S_{0}=0$.
  \For{$i=1:K$}\\
   Let $S_{i}=S_{i-1}+\frac{1}{\sigma_{d(i)}}$ and $\mu^{*}=\frac{i}{S_i+P}-\frac{V}{\tau  W_{k}(B_{l,k}-B_{r,k})}$.
        If $\mu^{*}\geq 0$,  $\frac{1}{\mu^{*}+\frac{V}{\tau  W_{k}(B_{l,k}-B_{r,k})}}-\frac{1}{\sigma_{d(i)}} \geq 0$, and $\frac{1}{\mu^{*}+\frac{V}{\tau  W_{k}(B_{l,k}-B_{r,k})}}-\frac{1}{\sigma_{d(i+1)}} \leq 0$, then 
        stop the loop, otherwise, move to the next iteration.
    \EndFor
\Statex  \textbf{step 5:} Let $\displaystyle\lambda^{*}_{i}=\max\left[0,  \frac{1}{\mu^{*}+\frac{V}{\tau  W_{k}(B_{l,k}-B_{r,k})}}-\frac{1}{\sigma_{i}}\right]$.
 \Statex  \textbf{step 6:} $\mathbf{Q}_{k}^{*} = \mathbf{U}^H\mathbf{\Lambda}^* \mathbf{U}$, with $\mathbf{\Lambda}^*$ diagonal matrix with entries $\lambda^{*}_{i}$
\end{algorithmic}
\end{algorithm}
\vspace{-.3cm}
   
\section{Numerical results}

In this section, numerical results are provided to assess the performance of our strategy. We consider a scenario with $N=6$ users aiming to offload their tasks to a MEH collocated at the AP serving the users. Each user is assigned an equal portion of the total bandwidth $B=1$ MHz, while the noise power spectral density is set to $N_0=-174$ dBm/Hz. The slot duration is set to $\tau=10$ ms. The arrival rate is $1$ Mbps with Poisson distribution, for all users. The maximum available CPU cycle frequency is $f_{\max}=4.5$ GHz and $J_k=500$ $\forall k$ (cf. \eqref{queue_evolution2}). 
All channels (cf. \eqref{chan}) are generated for a typical mmWave operating frequency, $f=28$ GHz, as in \cite{simR}, with: $K=4$, $N_a=4$, and $M=64$. The maximum transmit power for a single user $k$ is set to $P_k^{\max}=100$ mW. 
In the sequel, for the sake of comparison, we consider both scenarios with and without the RIS. Also, we assume two different degrees of channel knowledge: i) \textbf{Alg. 1:} instantaneous knowledge of $\beta_{k,r}(t)$ and $\beta_{k,a}(t)$ (cf. \ref{chan}) is assumed, and Alg. \ref{alg:cap} is used for radio resource allocation; ii) \textbf{Alg. 1, statistical}:  only a statistical knowledge of the blockage, i.e. the blocking probabilities $p_{k,a}$ and $p_{k,r}$ is assumed. In this case, Algorithm \ref{alg:cap} is used, but $\beta_{k,r}(t)$ and $\beta_{k,a}(t)$ are replaced by $p_{k,a}$ and $p_{k,r}$ in (\ref{chan}), for the optimization. Obviously, the data rate experienced by  each user is computed with the true channel in \eqref{chan}. 
Furthermore, for all cases, we consider also the case in which the RIS phase shifts are randomly selected; whereas, for \textbf{Alg. 1}, we also consider the case in which, after step 1.1 of Algorithm \ref{alg:cap}, the RIS phase shifts are quantized with $2$ bits, which is a practical constraint of RIS implementation \cite{14}. 

\begin{figure}[htb!]
    \centering
   \includegraphics[width=\columnwidth]{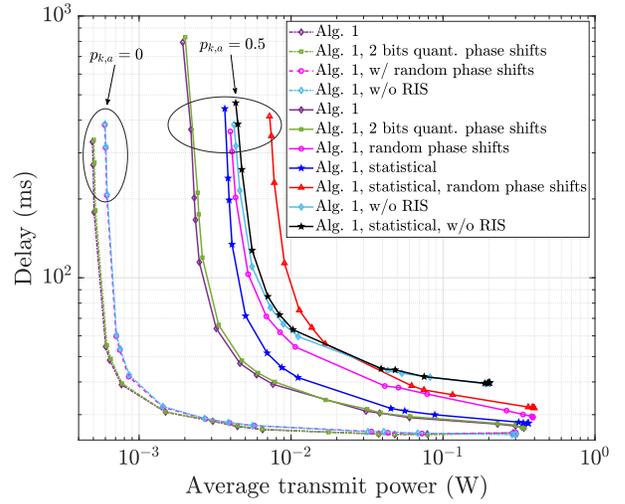}

    \caption{Delay-power trade-off}
    \label{fig:bp1}
\end{figure}

\begin{figure}[htb!]
    \centering
   \includegraphics[width=\columnwidth]{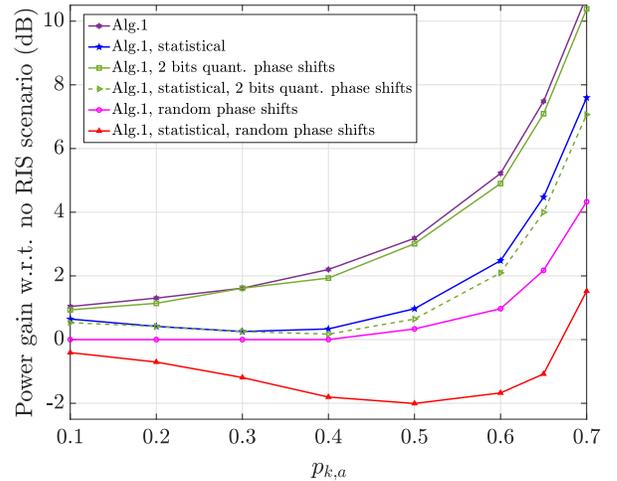}
    \caption{Avg. transmit power vs. AP blocking probability}
    \label{fig:trade_off_blockage}
\end{figure}
As a first result, in Fig. \ref{fig:bp1}, we show the trade-off between average E2E delay and transmit power, for different blocking probabilities $p_{k,a}$ on the direct link. We specifically plot the results for $p_{k,a}=0$ and $p_{k,a}=0.5$\footnote{For $p_{k,a}=0$, $\beta_{k,a}(t)=1,\forall k,t$, so that only \textbf{Alg. 1} is shown}, obtained by increasing the Lyapunov trade-off parameter $V$ from right to left. 
For all curves we can notice how, by increasing $V$, the system average transmit power  decreases while the average service delay increases. Also, all scenarios with the optimized RIS outperform the scenario without the RIS, although with negligible gain for the case without blocking. This suggests that the benefits of the RIS are more significant in case of high blocking probability of the direct link. Also, the \textbf{Alg. 1, statistical} strategy performs better than the non RIS case, except for the random phase shifts case. Finally, for \textbf{Alg. 1}, it can be noticed that a 2-bit quantization of RIS phase shifts yields parallel performance as the ideal case (i.e., continuous) with a very negligible gap, thus suggesting that our method can be exploited also for practical RIS optimization.\\ 
To further highlight the previously mentioned remarks, we illustrate, through Fig. \ref{fig:trade_off_blockage}, the gain in terms of average transmit power of each strategy with respect to the non RIS case, as a function of the direct link blocking\footnote{Problem \eqref{lg_eq} is not feasible for $p_{k,a}> 0.7$ without the RIS}, for a fixed E2E delay bound of $150$ ms, obtained by tuning the trade-off parameter $V$. As expected, as the blocking probability increases, the gain notably increases with the \textbf{Alg. 1} strategy, (up to 10 dB for $p_{k,a}=0.7$), also with quantized phases. Conversely, 
the gain of \textbf{Alg. 1, statistical} is visible only for higher blocking probabilities, due to the fact that, in this case, the channel knowledge is well-matched to the real channel states. Eventually, this implies that unreliable blocking knowledge is critical for the performance.
Optimizing the RIS through step $1.1$ of Algorithm \ref{alg:cap} leads to a better exploitation of the indirect path. 
However, for the random phase case, it can be noticed that no gain is achieved (less than 2 dB in the best case), which is obvious since we have no control on the RIS. More specifically, it can be noticed that for $p_{k,a}$ around $0.5$, the channel knowledge is completely mismatched. Instead, at high blocking probabilities $p_{k,a} \geq 0.6$, the mismatch is reduced and higher gain is achieved. 
Overall, we can conclude that the use of an RIS is prominent to satisfy a reliable MEC-based task offloading  in  case  of  bad  conditions of  the  direct  link, i.e., higher $p_{k,a}$, for all strategies. More specifically, the benefit of the RIS starts to be noticed with a lower $p_{k,a}$ for the best strategy, while it becomes more visible with higher $p_{k,a}$ for the worst strategy.

\section{Conclusion}
In this work, we have explored the effectiveness of using an RIS to counteract the  blocking (e.g. unreliability) when increasing communication frequency, in the case of computation offloading services.  To this end, we considered  a  blocking aware framework through which we investigated the dynamic joint optimization of computing  resources  and  RIS-aided multi-user MIMO  communication parameters. Then, for dynamic configuration, we applied Lyapunov optimization tools to transform a complex long-term optimization problem into a per-slot deterministic problem that requires only instantaneous observations of the context parameters and properly defined state variables. Numerical results show the inherent gain of empowering MEC with RISs, while assuming different degrees of channel knowledge along with different scenarios and blocking conditions.

\end{document}